\documentclass[twocolumn]{revtex4-2}
\usepackage{graphicx}
\usepackage{bm}
\usepackage{multirow}
\usepackage{array}
\usepackage{graphicx}
\usepackage{amsmath}
\usepackage{txfonts}
\usepackage{mathrsfs}
\usepackage{subfigure}
\usepackage{amssymb}
\usepackage{caption}
\usepackage{float}
\usepackage{booktabs}
\usepackage{diagbox}
\usepackage{setspace}
\usepackage{enumerate}
\usepackage{color}
\usepackage{tcolorbox}
\usepackage{inconsolata}
\definecolor{mygreen}{cmyk}{1,0.02,0.,0.42}        
\definecolor{myblue}{cmyk}{0.38,0.18,0.,0.46}

\usepackage{slashed}
\usepackage{lipsum}
\usepackage{hyperref}
\usepackage{url}
\hypersetup{hypertex=true,
	colorlinks=true,
	linkcolor=mygreen,
	anchorcolor=red,
	citecolor=blue}
\begin{document}
\title{Do we need dense matter equation of state in curved spacetime for neutron stars?}
\author{Jianing Li}
\author{Tao Guo} 
\author{Jiaxing Zhao} 
\author{Lianyi He}
\address{Department of Physics, Tsinghua University, Beijing 100084, China}

\date{\today}%

\begin{abstract}
Neutron stars are regarded as natural laboratories for the study of dense strong interaction matter. The equation of state (EoS) of dense matter computed in flat spacetime is used to predict the structure of neutron stars by solving the Tolman-Oppenheimer-Volkoff (TOV) equation.  Recently, it has been reported that the curved spacetime effect or specifically gravitational time dilation effect on the EoS of dense matter leads to a significant increase of the maximum mass limit of neutron stars [\href{https://journals.aps.org/prd/abstract/10.1103/PhysRevD.104.123005}{Phys. Rev. D \textbf{104}, 123005 (2021)} and \href{https://iopscience.iop.org/article/10.1088/1475-7516/2021/02/026}{J. Cosmol. Astropart. Phys. 02 (2021) 026}]. However, in this work, we show that to study the hydrostatic equilibrium of dense matter within the 
framework of general relativity and relativistic fluid dynamics, the grand canonical EoS of dense matter, $p(T,\mu)$, should be the same as that computed in flat spacetime, otherwise it is not consistent 
with local thermodynamic relations and energy-momentum conservation of the fluid. The gravitation influences the pressure $p$ only through enhancing the temperature $T$ and the chemical potential 
$\mu$, known as Tolman's law and Klein's law.  We rewrite the TOV equation as an alternative version so that the grand canonical EoS computed by using field theoretical methods can be used as a direct input. This may provide a tool to study the grand canonical EoS of dense matter via deep learning.
\end{abstract}

\maketitle

\section{Introduction}\label{Sec1}

Neutron stars could be one kind of the densest objects in our universe. Their masses are estimated to be between the Chandrasekhar limit $1.4~M_\odot$ and $2.16~M_\odot$~\cite{glendenning2012compact,Mazzali:2007et,Rezzolla:2017aly}, with $M_\odot$ the solar mass. The most massive neutron star that has ever been observed so far
is the one in the binary system PSR J0740+6620, consisting of a neutron star and a white dwarf~\cite{NANOGrav:2019jur}. The mass of the neutron star in this binary system is reported to be $2.08\pm{0.07}~M_\odot$, close to the upper mass limit~\cite{Fonseca:2021wxt}. Neutron stars have been regarded as natural laboratories for the study of the many-body physics of dense strong interaction matter. Neutron stars are usually thought to be composed of neutron matter at a few times of the nuclear saturation density, with a small amount of protons and leptons to ensure charge neutrality and beta equilibrium. It is also conjectured that with increasing matter density, deconfined quark matter may emerge in the core of neutron stars~\cite{Annala:2019puf}. The strangeness component may also appear, such as hyperons and even strange quark matter~\cite{Bombaci:2020vgw,Ellis:1990qq,Masuda:2012kf}.  

While numerous observations for neutron stars have been accumulated, plenty of unclear issues still remain. For example, the emergence of hyperons seems inevitable if the matter becomes sufficiently dense. However, this will soften the equation of state (EoS) and make the largest mass predicted by theory smaller than the Chandrasekhar limit. This is the so-called hyperon puzzle~\cite{Bombaci:2016xzl}, which may originate from the obscurity of the interactions in the many-hyperon system~\cite{Masuda:2015kha,Jiang:2012hy}. It is also debated whether there exists 
quark matter in neutron stars~\cite{Mazzali:2007et,Somasundaram:2021clp}. This is related to the theoretical issue of the transition from hadronic matter to quark matter. To solve these issues, one of the top priorities is to compute the accurate EoS of dense strong interaction matter. 

On the theoretical side, plenty of phenomenological models for the nuclear force ~\cite{Nishizaki:2001in,Nishizaki:2002ih,Baldo:1999rq,Machleidt:2000ge} have been used to predict the EoS of dense nuclear matter.  Quantum field theory is also a powerful tool to calculate the EoS of relativistic dense matter~\cite{kapusta_gale_2006}. On the other hand, to predict the structure of (static) neutron stars, we solve the Tolman-Oppenheimer-Volkoff (TOV) equation~\cite{Oppenheimer:1939ne,Tolman:1939jz}, with the dense matter EoS as an input. However, the EoS of dense matter are usually computed by using quantum many-body theory in flat spacetime, and the possible curved spacetime effect induced by strong gravitation in neutron stars is not taken into account at all.  
Therefore, it seems discordant to put the dense matter EoS computed in flat spacetime into the TOV equation. A question naturally comes into being: Can we use the dense matter EoS computed in flat spacetime to study neutron stars? If curved spacetime effect really influences the dense matter EoS, it would increase the complexity of the study of dense matter from neutron stars.

Recently, some works have reported that the dense matter EoS computed by using quantum field theory in curved spacetime would make a big difference~\cite{Hossain:2020mvn,Hossain:2021qyg}. 
They found that the gravitational time dilation effect leads to a significant increase of the maximum mass of neutron stars. In this work, however, we clarify that to study the hydrostatic equilibrium of 
dense matter within the framework of general relativity and relativistic fluid dynamics, the grand canonical EoS of dense matter, i.e., the pressure $p$ as a function of the temperature $T$ and the chemical potential $\mu$, $p(T,\mu)$, should be the same as that computed in flat spacetime. We show that this is a requirement from the local thermodynamic relations and the conservation of the energy and momentum of the relativistic fluid. The gravitation influences the pressure $p$ only through enhancing the temperature $T$ and the chemical potential $\mu$, known as Tolman's law~\cite{Tolman:1930zza,Tolman:1930ona} and Klein's law~\cite{klein1949thermodynamical}.  Hence the theoretical framework of TOV equation and relativistic fluid dynamics with an EoS determined from flat spacetime is self-consistent.

The paper is organized as follows. In Sec. \ref{Sec2} we prove that the dense matter EoS used to study the hydrostatic equilibrium in a static and spherical star should be the same as that in the flat spacetime. We generalize the proof to general static spacetime in Sec. \ref{Sec3}. In Sec. \ref{Sec4}, we convert the TOV equation into a grand canonical version so that the grand canonical EoS can be used as a direct input. We demonstrate the solution and visualize the gravitational effect on the baryon chemical potential by using the Walecka model. We summarize in Sec. \ref{Sec5}. The nature units
$c=\hbar=k_{\rm B}=1$ are used throughout.

\section{The grand canonical EoS in local thermal equilibrium}\label{Sec2}
Consider isolated dense matter in hydrostatic equilibrium. A curved spacetime is created according to general relativity and we assume that it is spherically symmetric. The line element $\text{d}s^2=g_{\mu\nu}\text{d}x^\mu \text{d}x^\nu$ can be written as
\begin{equation}
\text{d}s^2=-e^{2\Phi\left(r\right)}\text{d}t^2+e^{2\Psi\left(r\right)}\text{d}r^2+r^2\text{d}\theta^2+r^2\sin^2\theta\text{d}\phi^2.
\end{equation}
The spacetime metric reads explicitly  
\begin{eqnarray}
&&g_{tt}=-e^{2\Phi\left(r\right)}, \ \ \ g_{rr}=e^{2\Psi\left(r\right)}, \nonumber\\
&&g_{\theta\theta}=r^2, \ \ \ \ g_{\phi\phi}=r^2\sin^2\theta, \nonumber\\
&&g_{\mu\nu}=0 \ \ \ \text{for} \ \ \mu\neq\nu.
\end{eqnarray}
The dense matter can be described by relativistic fluid dynamics. For a relativistic fluid, the energy-momentum tensor can be written as
\begin{equation}
T_{\mu\nu}=pg_{\mu\nu}+\left(p+\varepsilon\right)U_\mu U_\nu+\pi_{\mu\nu},
\end{equation}
with $p$ the isotropic pressure and $\varepsilon$ the proper energy density. For hydrostatic equilibrium, the transport terms in $\pi_{\mu\nu}$ do not contribute and can be neglected from now on. The velocity four-vector $U^\mu$ is defined so that $g^{\mu\nu}U_\mu U_\nu=-1$. 
Since the fluid is at rest, we take
\begin{equation}
U_t=\sqrt{-g_{tt}}=e^{\Phi},\ \ \ U_r=U_\theta=U_\phi=0.
\end{equation}

The matter profile and the spacetime metric can be determined by solving Einstein's field equation $G_{\mu\nu}=8\pi GT_{\mu\nu}$, with $G$ the gravitational constant. Computing the Einstein tensor $G_{\mu\nu}$ we obtain a number of equations~\cite{Weinberg:1972kfs,carroll_2019}. The $tt$-component gives
\begin{equation}
e^{2\Psi(r)}=\left(1-\frac{2Gm}{r}\right)^{-1},
\end{equation} 
where 
\begin{equation}
m(r)\equiv\int_0^r4\pi r^2\varepsilon \text{d}r
\end{equation}
can be interpreted as the total mass contained inside radius $r$. The $rr$-component gives
\begin{equation}
\frac{\text{d}\Phi\left(r\right)}{\text{d}r}=\frac{e^{2\Psi(r)}}{r^2}G\left(m+4\pi r^3p\right).
\end{equation}
The third equation can be derived from the $\theta\theta$-component or $\phi\phi$-component. However, it is convenient to use continuity equation of the energy-momentum tensor, 
$\nabla_\mu T^{\mu\nu}=0$, which is guaranteed by Einstein's field equation. The only nontrivial equation is given by the $\nu=1$ (or $\nu=r$) component. A direct calculation gives
\begin{equation}
\nabla_\mu T^{\mu1}=e^{-2\Psi\left(r\right)}\left[\frac{\text{d}p}{\text{d}r}+(p+\varepsilon)\frac{\text{d}\Phi\left(r\right)}{\text{d}r}\right].
\end{equation}

Summarizing the above results, we finally arrive at the famous Tolman-Oppenheimer-Volkoff equation
\begin{equation}\label{TOV}
\frac{\text{d}p}{\text{d}r}=-\frac{G(p+\varepsilon)(m+4\pi r^3p)}{r^2\left(1-\frac{2Gm}{r}\right)}.
\end{equation}
This equation is normally solved by using the dense matter EoS of the form $p=p(\varepsilon)$ as an input. However, in this form, it is not quite clear whether and how the gravitational effect on the EoS should be
taken into account. Actually, we normally use the dense matter EoS determined in flat spacetime or on Earth.

On the other hand, theorists are good at computing the grand canonical EoS $p=p(T,\mu)$ by using finite temperature field theory in flat spacetime~\cite{kapusta_gale_2006}.  Some recent works have tried to compute the grand canonical EoS based on the statistic mechanics of quantum fields in curved spacetime~\cite{Hossain:2020mvn,Hossain:2021qyg}. Within their approach, the gravitational time dilation effect explicitly influences the grand canonical EoS, i.e., $p(T,\mu)$ is different at different position in the gravitational field. This leads to a significant increase of the maximum mass limit of neutron stars, in contrast to the previous predictions based on the dense matter EoS determined in flat spacetime. In the following, we will discuss how the local thermodynamic relations and the energy-momentum conservation in fluid dynamics (as guaranteed by Einstein's field equation) constrain the grand canonical EoS $p(T,\mu)$, or, what kind of EoS $p(T,\mu)$ is compatible with the TOV equation and local thermal equilibrium.

The dense matter described by relativistic fluid dynamics is composed of many fluid elements sufficiently small. On the other hand, each small fluid element should contain sufficiently large degrees of freedom so that the thermodynamic limit and local thermal equilibrium can be reached.  In curved spacetime, each fluid element is described by local thermodynamic variables in the local rest frame of the fluid element. According to the equivalence principle, these local thermodynamic variables should obey the fundamental laws of thermodynamics~\cite{Tolman:1930zza,Tolman:1930ona,klein1949thermodynamical,Aydemir:2021dan,Lima:2019brf,Kim:2021kou}. 

Assuming that the relativistic fluid may carry several conserved charges, we introduce the corresponding chemical potentials $\mu_1,\mu_2,...$, denoted by $\{\mu_i\}$ for convenience. The grand canonical EoS of a fluid element at the position $(r,\theta,\phi)$ can be formally expressed as
\begin{equation}
p=p(T,\{\mu_i\}; r).
\end{equation}
Here we first assume that the EoS may be different at different positions in curved spacetime, i.e., the gravitation caused a direct influence on the EoS. Because of the isotropy, the explicit position dependence can be realized only through the radius $r$. Note that $T$ and $\mu_i$ are the local temperature and chemical potentials of the fluid element located at the position $(r,\theta,\phi)$, i.e.,
\begin{equation}
T=T(r),\ \ \ \ \mu_i=\mu_i(r).
\end{equation}
 If local thermal equilibrium is reached, the local thermodynamic quantities should satisfy the fundamental thermodynamic relation
\begin{equation}\label{FTR}
\varepsilon=Ts+\sum_{i}\mu_i n_i-p,
\end{equation}
where the entropy density $s$ and the number density $n_i$ can be evaluated from the EoS,
\begin{eqnarray}\label{GrandR}
s=\frac{\partial p(T,\{\mu_i\}; r)}{\partial T},\ \ \ \ \ n_i=\frac{\partial p(T,\{\mu_i\}; r)}{\partial \mu_i}.
\end{eqnarray}

For a relativistic fluid in hydrostatic equilibrium, the conservation of the energy and momentum, $\nabla_\mu T^{\mu\nu}=0$, gives
\begin{equation}\label{EMC}
\frac{\text{d}p}{\text{d}r}=-(p+\varepsilon)\frac{\text{d}\Phi}{\text{d}r}.
\end{equation}
Using the fundamental thermodynamic relation (\ref{FTR}), we arrive at
\begin{equation}
\frac{\text{d}p}{\text{d}r}=-\left(Ts+\sum_i\mu_in_i\right)\frac{\text{d}\Phi}{\text{d}r}.
\end{equation}
Further using the thermodynamic relation (\ref{GrandR}), we obtain a functional equation
\begin{widetext}
\begin{equation}
\frac{\text{d}p}{\text{d}r}=\frac{\partial p(T,\{\mu_i\}; r)}{\partial T}\left(-T\frac{\text{d}\Phi}{\text{d}r}\right)+\sum_{i}\frac{\partial p(T,\{\mu_i\}; r)}{\partial \mu_i}\left(-\mu_i \frac{\text{d}\Phi}{\text{d}r}\right),
\end{equation}
which is valid at arbitrary radius $r$. On the other hand, the standard chain rule gives
\begin{equation}
\frac{\text{d}p}{\text{d}r}=\frac{\partial p(T,\{\mu_i\}; r)}{\partial r}
+\frac{\partial p(T,\{\mu_i\}; r)}{\partial T}\frac{\text{d}T}{\text{d}r}
+\sum_{i}\frac{\partial p(T,\{\mu_i\}; r)}{\partial \mu_i}\frac{\text{d}\mu_i}{\text{d}r}.
\end{equation} 
\end{widetext}
Comparing the above two equations for arbitrary radius $r$, we find
\begin{eqnarray}
\frac{\partial p(T,\{\mu_i\}; r)}{\partial r}=0.
\end{eqnarray} 
Thus we conclude that \emph{the grand canonical EoS does not depend explicitly on the position}. Meanwhile, we obtain other two relations
\begin{eqnarray}\label{TMU}
&&\frac{\text{d}T}{\text{d}r}=-T\frac{\text{d}\Phi}{\text{d}r},\nonumber\\
&&\frac{\text{d}\mu_i}{\text{d}r}=-\mu_i\frac{\text{d}\Phi}{\text{d}r}.
\end{eqnarray} 
The solution of these two equations can be expressed as
\begin{eqnarray}\label{Tolman}
T(r)&=&T_\infty e^{-\Phi(r)},\nonumber\\
\mu_i(r)&=&\mu_{i}^\infty e^{-\Phi(r)}.
\end{eqnarray} 
They are nothing but the Tolman's law for the temperature~\cite{Tolman:1930zza,Tolman:1930ona} and the Klein's law for the chemical potentials~\cite{klein1949thermodynamical} in a static gravitational field. Here the constants $T_\infty$ and $\mu_{i}^\infty$ are interpreted as the temperature and chemical potentials measured by an observer at infinity ($r\rightarrow\infty$).  The above laws also guarantee that the fugacities $z_i=\exp(\mu_i/T)$ are position independent, as required by vanishing heat flow and diffusion for a system in local thermal equilibrium~\cite{Israel:1976tn}.

Summarizing the above results, we conclude that the grand canonical EoS should be uniform in a curved spacetime created by dense matter in hydrostatic equilibrium, i.e.,
\begin{equation}
p=p(T,\{\mu_i\}).
\end{equation}
The gravitation influences the pressure only through the redshift of the temperature and chemical potentials, i.e., the Tolman's law and the Klein's law. Since the spacetime is asymptotically flat,
we can determine the grand canonical EoS $p(T,\{\mu_i\})$ at $r\rightarrow\infty$, that is, \emph{the grand canonical EoS can be essentially determined in flat spacetime}.

In previous works~\cite{Hossain:2020mvn,Hossain:2021qyg}, the gravitational effect is attributed to the gravitational potential $\Phi(r)$, i.e., the gravitational time dilation.  Since $\Phi(r)$ is a single-valued function of $r$, it is equivalent to assume that the EoS may depend explicitly on $\Phi$, i.e.
\begin{equation}
p=p(T,\{\mu_i\}; \Phi).
\end{equation}
The conservation of energy and momentum, Eq. (\ref{EMC}), can be rewritten as
\begin{equation}
\frac{\text{d}p}{\text{d}\Phi}=-p-\varepsilon.
\end{equation}
Using the thermodynamic relations, we obtain
\begin{equation}
\frac{\text{d}p}{\text{d}\Phi}=-T\frac{\partial p(T,\{\mu_i\}; \Phi)}{\partial T}-\sum_{i}\mu_i \frac{\partial p(T,\{\mu_i\}; \Phi)}{\partial \mu_i}.
\end{equation}
On the other hand, the standard chain rule gives
\begin{eqnarray}
\frac{\text{d}p}{\text{d}\Phi}&=&\frac{\partial p(T,\{\mu_i\}; \Phi)}{\partial T}\frac{\text{d}T}{\text{d}\Phi}
+\sum_{i}\frac{\partial p(T,\{\mu_i\}; \Phi)}{\partial \mu_i}\frac{\text{d}\mu_i}{\text{d}\Phi}\nonumber\\
&+&\frac{\partial p(T,\{\mu_i\}; \Phi)}{\partial \Phi}.
\end{eqnarray} 
Therefore, we have the following identities
\begin{eqnarray}
&&\frac{\partial p(T,\{\mu_i\}; \Phi)}{\partial \Phi}=0,\nonumber\\
&&\frac{\text{d}T}{\text{d}\Phi}=-T,\ \ \ \ \ \ \frac{\text{d}\mu_i}{\text{d}\Phi}=-\mu_i.
\end{eqnarray} 
The first identity indicates that \emph{the grand canonical EoS does not depend explicitly on $\Phi$}.  The second and the third equations give the same results as in Eq. (\ref{Tolman}).
We note that the grand canonical EoS computed in previous works~\cite{Hossain:2020mvn,Hossain:2021qyg}, which shows an explicit dependence on the gravitational potential $\Phi$, 
\emph{is not compatible with the TOV equation and local thermodynamic relations}.

\section{Generalization to arbitrary static spacetime}\label{Sec3}
Even though the configuration of dense matter in hydrostatic equilibrium is normally spherically symmetric,  the results in Sec. \ref{Sec2} can be generalized to
arbitrary static spacetime. Consider a general static curved spacetime. The line element $\text{d}s^2=g_{\mu\nu}\text{d}x^\mu \text{d}x^\nu$ can be expressed as ($x^0\equiv t$)
\begin{equation}
\text{d}s^2=g_{00}\text{d}t^2+g_{ij}\text{d}x^i\text{d}x^j.
\end{equation}
The metric functions $g_{00}$ and $g_{ij}$ are independent of time but depend in an arbitrary way of the spatial coordinates $x^k$ ($k=1,2,3$).
For hydrostatic equilibrium, we take
\begin{equation}
U_0=\sqrt{-g_{00}},\ \ \ U_1=U_2=U_3=0.
\end{equation}
 
We also consider the conservation of the energy and momentum, $\nabla_\mu T^{\mu\nu}=0$. The $\nu=k$ ($k=1,2,3$) component gives
\begin{equation}
\frac{\text{d}p}{\text{d}x^k}=-(p+\varepsilon)\frac{\text{d}\ln\sqrt{-g_{00}}}{\text{d}x^k}.
\end{equation}
Here we use $\text{d}/\text{d}x^k$ to denote the derivative with respect to the spatial coordinates, so that it can be distinguished from that with respect to the temperature and chemical potentials. 
The grand canonical EoS of a fluid element at the position $(x^1,x^2,x^3)$ can be formally expressed as
\begin{equation}
p=p(T,\{\mu_i\}; \{x^k\}).
\end{equation}
Using the local thermodynamic relations, we obtain
\begin{widetext}
\begin{equation}
\frac{\text{d}p}{\text{d}x^k}=\frac{\partial p(T,\{\mu_i\}; \{x^k\})}{\partial T}\left(-T\frac{\text{d}\ln\sqrt{-g_{00}}}{\text{d}x^k}\right)+\sum_{i}\frac{\partial p(T,\{\mu_i\}; \{x^k\})}{\partial \mu_i}
\left(-\mu_i \frac{\text{d}\ln\sqrt{-g_{00}}}{\text{d}x^k}\right).
\end{equation}
On the other hand, the standard chain rule gives
\begin{equation}
\frac{\text{d}p}{\text{d}x^k}=\frac{\partial p(T,\{\mu_i\}; \{x^k\})}{\partial x^k}+\frac{\partial p(T,\{\mu_i\}; \{x^k\})}{\partial T}\frac{\text{d}T}{\text{d}x^k}+\sum_{i}\frac{\partial p(T,\{\mu_i\}; \{x^k\})}{\partial \mu_i}
\frac{\text{d}\mu_i}{\text{d}x^k}.
\end{equation}
\end{widetext}
Again, comparison of the two results gives the following identities
\begin{eqnarray}
	\label{33}
&&\frac{\partial p(T,\{\mu_i\}; \{x^k\})}{\partial x^k}=0,\nonumber\\
&&\frac{\text{d}T}{\text{d}x^k}=-T\frac{\text{d}\ln\sqrt{-g_{00}}}{\text{d}x^k},\nonumber\\
&&\frac{\text{d}\mu_i}{\text{d}x^k}=-\mu_i\frac{\text{d}\ln\sqrt{-g_{00}}}{\text{d}x^k}.
\end{eqnarray} 
The first identity indicates that \emph{the grand canonical EoS should not depend on the spatial position explicitly in a general static spacetime}. The second and the third identities give the Tolman's law and the Klein's law in a general static spacetime,
\begin{eqnarray}
T(\{x^k\})&=&\frac{T_\infty}{\sqrt{-g_{00}(\{x^k\})}},\nonumber\\
\mu_i(\{x^k\})&=&\frac{\mu_{i}^\infty}{\sqrt{-g_{00}(\{x^k\})}}.
\end{eqnarray} 
If the spacetime is asymptotically flat as normally satisfied, this means that the grand canonical EoS $p(T,\{\mu_i\})$ is the same as that determined in flat spacetime. Otherwise it is not consistent
with local thermodynamic relations. 

In fact, applying EoS in flat spacetime as an input is correct in the view of general relativity. As the process shown in Sec.~\ref{Sec2}, the TOV equation is a tensor equation derived from the conservation law of energy-momentum tensor. The pressure in the TOV equation is the one appearing in the energy-momentum tensor. The equivalence principle states that any physical laws expressed in the form of tensor equations in special relativity holds in the same form in general relativity~\cite{Weinberg:1972kfs,carroll_2019}. Therefore, the EoS imported to the TOV equation does not need any general-relativistic correction. Such correction is ought to be inherited by the covariant derivative terms instead. Although this validity can be deduced based on general relativity, Eq.~(\ref{33}) can only derived from equilibrium thermodynamic.

\section{Neutron stars from $\boldsymbol{p}\left(\boldsymbol{\mu}\right)$ EoS}\label{Sec4}
The TOV equation (\ref{TOV}) is convenient to solve if the EoS of the form $p=p(\varepsilon)$ is known. However, as field theoretical methods normally determine the grand canonical EoS $p=p(T,\{\mu_i\})$ directly, it is more convenient to use Eq. (\ref{TMU}) and arrive at an alternative version of the TOV equation. For a cold star, we can set $T=0$ and obtain
\begin{equation}\label{TOV2}
	\left\{
	\begin{split}
		&\frac{\text{d}\mu_i}{\text{d}r}=-\mu_i\frac{\text{d}\Phi}{\text{d}r},\\
		&\frac{\text{d}\Phi}{\text{d}r}=\frac{G(m+4\pi r^3p)}{r^2\left(1-\frac{2Gm}{r}\right)}.
	\end{split}
	\right.
\end{equation}
If we are not interested in the gravitational potential $\Phi$, we can write
\begin{equation}\label{TOV3}
\frac{\text{d}\mu_i}{\text{d}r}=-\frac{G\mu_i(m+4\pi r^3p)}{r^2\left(1-\frac{2Gm}{r}\right)}.
\end{equation}
This version can be conveniently solved if the grand canonical EoS $p=p(\{\mu_i\})$ at zero temperature is known. The energy density in the mass function $m(r)$ can be expressed as
\begin{equation}
\varepsilon(\{\mu_i\})=\sum_i\mu_i\frac{\partial p(\{\mu_i\})}{\partial \mu_i}-p(\{\mu_i\}).
\end{equation}
For neutron stars, there is only one chemical potential, the baryon chemical potential $\mu_{\rm B}$, serves as the thermodynamic variable in the EoS.

In the following, we adopt the Walecka model~\cite{Walecka:1974qa} to describe the dense matter in neutron stars and demonstrate the solution of the TOV equation (\ref{TOV2}) with the grand canonical EoS. As clarified in Sec. \ref{Sec2}, we only need to calculate the EoS from the model in flat spacetime. The Lagrangian density of the Walecka model is given by
\begin{equation}
	\begin{split}
		\mathcal{L}_{\rm W}=& \sum_{\rm N=n,p} \bar{\psi}_{\rm N}\left(i \gamma^\mu\partial_\mu-m_{\rm N}+g_{\sigma } \sigma-g_{\omega } \gamma^\mu\omega_\mu\right) \psi_{\rm N} \\
		&+\frac{1}{2}\left(\partial_{\mu} \sigma \partial^{\mu} \sigma-m_{\sigma}^{2} \sigma^{2}\right)-U(\sigma) \\
		&-\frac{1}{4} F^{\mu \nu} F_{\mu \nu}+\frac{1}{2} m_{\omega}^{2} \omega^{\mu} \omega_{\mu},
	\end{split}
\end{equation}
where $F_{\mu\nu}=\partial_\mu \omega_\nu-\partial_\nu \omega_\mu$ and $\psi_{\rm N}$ (${\rm N=n,p}$) denote the nucleon fields with mass $m_{\rm N}$. In the present model, isospin symmetry is assumed for the sake of simplicity. The scalar $\sigma$ meson with mass $m_\sigma$ and the vector $\omega$ meson with mass $m_\omega$ are introduced to describe the long-range attraction and the short-range repulsion of the nuclear force. This is realized by the two meson-nucleon coupling terms with coupling constants $g_\sigma$ and $g_\omega$. We add a phenomenological potential
$U(\sigma)=\frac{1}{3} b m_{\rm N}\left(g_{\sigma} \sigma\right)^{3}+\frac{1}{4} c\left(g_{\sigma} \sigma\right)^{4}$ to fit the empirically known properties of nuclear matter~\cite{kapusta_gale_2006}.

To describe the neutron star matter, electrons and even muons should be introduced to guarantee $\beta$-equilibrium and charge neutrality.  We thus add a term for leptons,
\begin{equation}
\mathcal{L}_{\rm lep}= \sum_{l={\rm e},\mu} \bar{\psi}_{l}\left(i \gamma^\mu\partial_\mu-m_l\right) \psi_{l}.
\end{equation}
The partition function of the model can be expressed in the imaginary-time path integral formalism,
\begin{eqnarray}
{\cal Z}&=& \int \prod_\alpha[d\psi_\alpha][d\bar{\psi}_\alpha][d\sigma][d\omega_\mu]\nonumber\\
&\times&\exp\left\{\int_0^\beta \text{d}\tau \int \text{d}^3{\bf x} \left[{\cal L}_{\rm W}+{\cal L}_{\rm lep}+\sum_\alpha\mu_\alpha\psi_\alpha^\dagger\psi_\alpha^{\phantom{\dag}}\right]\right\},\nonumber\\
\end{eqnarray}
where $\alpha$ denotes the fermion species, i.e., $\alpha={\rm n}, {\rm p}, \text{e},\mu$, and $\beta=1/T$. The four chemical potentials $\mu_\alpha$ are not independent. They are constrained by $\beta$-equilibrium. It is 
equivalent to introduce only the baryon number $\text{B}=\sum_{\rm N=n,p}\psi^\dagger_{\rm N}\psi^{\phantom{\dag}}_{\rm N}$ and electric charge $\text{Q}=\sum_{\rm a=p,e,\mu}\psi^\dagger_{\rm a}\psi^{\phantom{\dag}}_{\rm a}$, which are conserved in the presence of weak interaction. Thus we can express the four chemical potentials $\mu_\alpha$ in terms of the
baryon chemical potential $\mu_{\rm B}$ and the electric chemical potential $\mu_{\rm Q}$ as 
\begin{equation}
\mu_{\rm n}=\mu_{\rm B},\quad\mu_{\rm p}=\mu_{\rm B}+\mu_{\rm Q},\quad\mu_{\text{e}}=\mu_{\mu}=\mu_{\rm Q}.
\end{equation}
For neutron stars mainly composed of neutrons, the electric chemical potential $\mu_{\rm Q}$ is negative.

The partition function and the thermodynamic quantities can be conveniently computed within the mean field approximation~\cite{kapusta_gale_2006}. At finite density, the nucleons act as sources in the equations of motion for the meson fields, which indicates that finite density generates nonzero expectation values for the scalar and vector meson fields. In the mean field approximation, the effective potential for the static and uniform meson fields $\sigma$ and $\omega_0$ is given by
\begin{eqnarray}
&&{\cal V}_{\rm eff}(\sigma,\omega_0;T,\mu_{\rm B},\mu_{\rm Q})\nonumber\\
&=&\frac{1}{2}m_\sigma^2\sigma^2+U(\sigma)-\frac{1}{2}m_\omega^2\omega_0^2\nonumber\\
&-&\frac{T}{V}\sum_{n,{\bf k}}\sum_{\alpha}\ln\det\left[{\cal S}_{0\alpha}^{-1}(ik_n,{\bf k})+\Sigma_\alpha(\sigma,\omega_0)\right],
\end{eqnarray}
where ${\cal S}_{\rm0\alpha}^{-1}=(ik_n+\mu_\alpha)\gamma^0-\mbox{\boldmath{$\gamma$}}\cdot{\bf k}-m_\alpha$ is the inverse of the thermal Green's function of free nucleons and leptons, with $k_n=(2n+1)\pi T$ ($n \in \mathbb{Z}$). The quantities $\Sigma_\alpha$ are defined as $\Sigma_{\rm n}=\Sigma_{\rm p}=g_{\sigma } \sigma-g_{\omega } \gamma^0\omega_0$ and 
$\Sigma_{\rm e}=\Sigma_{\mu}=0$.
The physical values of the classical meson fields $\bar{\sigma}$ and $\bar{\omega}_0$ are determined by minimizing of the effective potential,
\begin{equation}
\frac{\partial {\cal V}_{\rm eff}(\sigma,\omega_0)}{\partial\sigma}=0,\ \ \ \frac{\partial {\cal V}_{\rm eff}(\sigma,\omega_0)}{\partial\omega_0}=0.
\end{equation}
These extreme equations determine $\bar{\sigma}$ and $\bar{\omega}_0$ as functions of the temperature and chemical potentials. As this is done, the thermodynamic quantities can be obtained.
The pressure is given by 
\begin{equation}
p(T,\mu_{\rm B},\mu_{\rm Q})=-{\cal V}_{\rm eff}(\bar{\sigma},\bar{\omega}_0;T,\mu_{\rm B},\mu_{\rm Q}).
\end{equation}

\begin{figure}[h]
	\centering
	{\includegraphics[width=0.5\textwidth]{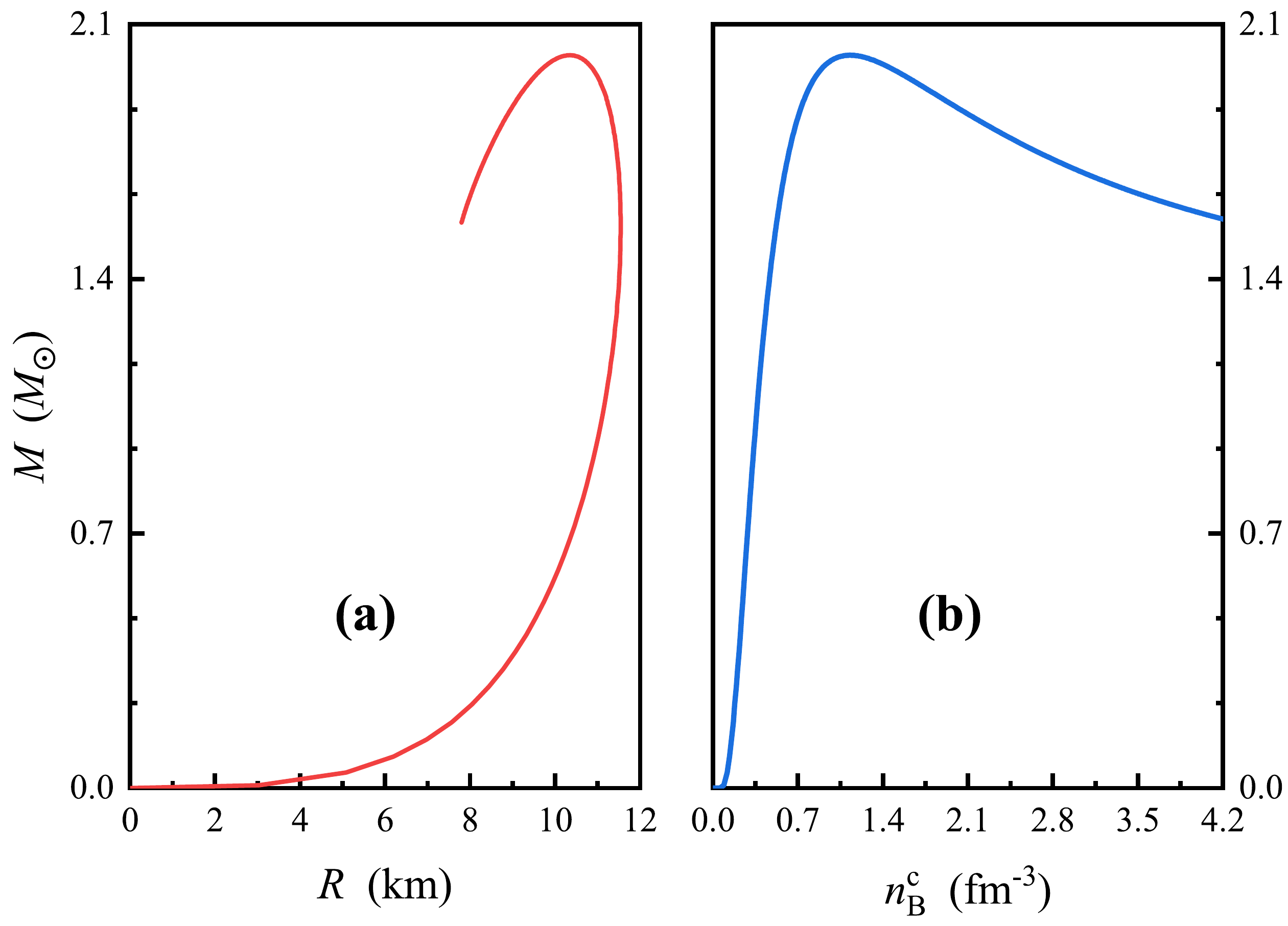}}
	\caption{The mass-radius relation of neutron stars (a) and the relation between the neutron star mass and the central baryon density (b) calculated from the Walecka model.}
	\label{NS mass}
\end{figure}

For neutron star matter, electric charge neutrality requires that the net electric charge density should vanish, that is
\begin{equation}
\frac{\partial p(T,\mu_{\rm B},\mu_{\rm Q})}{\partial\mu_{\rm Q}}=0.
\end{equation}
Hence the electric chemical potential $\mu_{\rm Q}$ is not an independent thermodynamic variable. We should solve the above equation to obtain $\mu_{\rm Q}=\mu_{\rm Q}(T,\mu_{\rm B})$. 
Therefore, only $T$ and $\mu_{\rm B}$ are independent thermodynamic variables and the grand canonical EoS takes the form $p=p(T,\mu_{\rm B})$. In static neutron stars, they should satisfy the
Tolman' law $T=T_\infty e^{-\Phi}$ and the Klein's law $\mu_{\rm B}=\mu_{\rm B}^\infty e^{-\Phi}$.

The temperature of a stable neutron star is typically low and we can set $T=0$. The zero temperature EoS $p(\mu_{\rm B})$ is evaluated in Appendix A. We use this EoS to solve the TOV equation (\ref{TOV2}) and visualize the gravitational effect on the baryon chemical potential 
$\mu_{\rm B}$. Briefly speaking,  starting with a given local baryon chemical potential $\mu_{\rm B}^\text{c}$ at the core ($r=0$), we obtain the profiles of the pressure and the baryon chemical potential from the TOV equation (\ref{TOV3}). The pressure decreases to zero at the surface, which determines the mass $M$ and radius $R$ of the neutron star and hence the mass-radius relation as displayed in Fig.~\ref{NS mass}.  The gravitational potential at the surface is then known as $\Phi_{\rm s}=\Phi(R)=\frac{1}{2}\ln(1-2GM/R)$. The baryon chemical potential satisfies the Klein's law $\mu_{\rm B}(r)=\mu_{\rm B}^\infty e^{-\Phi(r)}$. The constant $\mu_{\rm B}^\infty$ can be determined as $\mu_{\rm B}^\infty=\mu_{\rm B}^{\rm s}e^{\Phi_{\rm s}}$, where 
$\mu_{\rm B}^{\rm s}$ is the baryon chemical potential at the surface. Note that $\mu_{\rm B}^{\rm s}$ is purely determined by the EoS, $p(\mu_{\rm B}^{\rm s})=0$, i.e., the critical chemical potential
that separates the vacuum and the matter phase.

\begin{figure}[h]
	\centering
	{\includegraphics[width=0.42\textwidth]{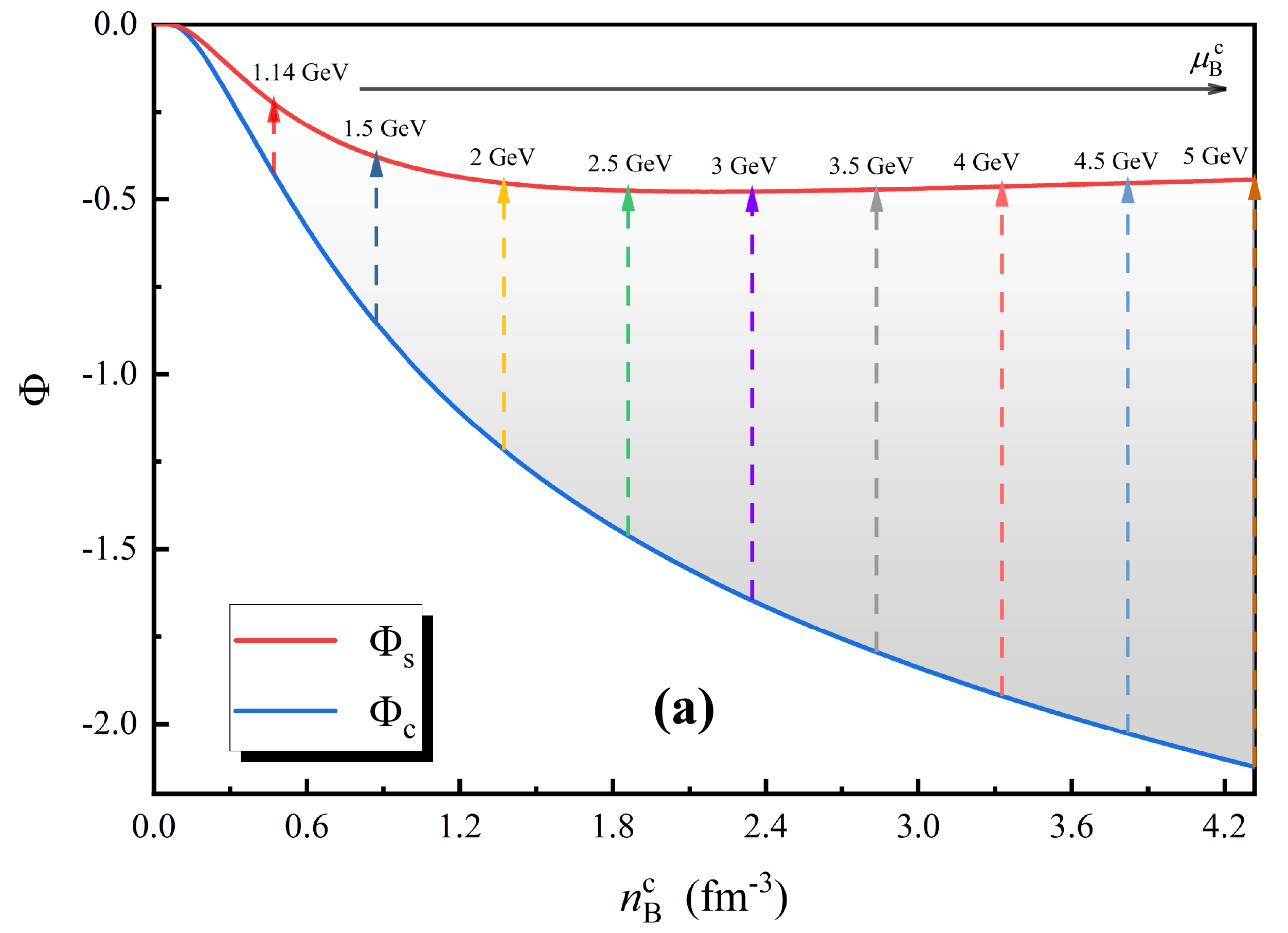}}
	{\includegraphics[width=0.4\textwidth]{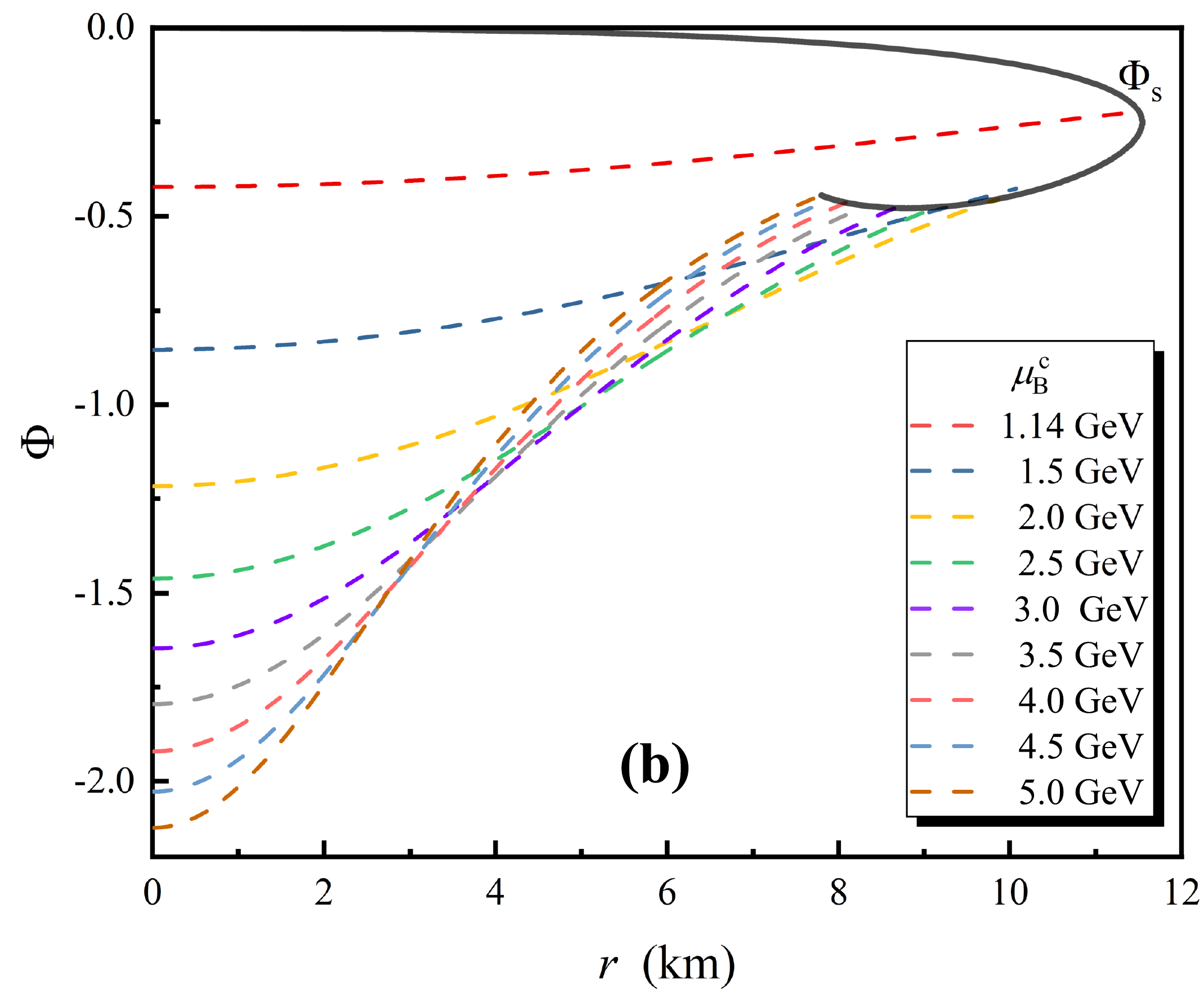}}
	\caption{The gravitational potential $\Phi$ in the interior of neutron stars computed from the Walecka model.  (a) $\Phi$ at the surface and at the core as functions of the central baryon density $n_{\rm B}^{\rm c}$. The colored dashed lines with arrows denote the corresponding baryon chemical potentials $\mu_{\rm B}^{\rm c}$ at the core. (b) The profile $\Phi(r)$ (thin dashed lines) in the interior of neutron stars with different central densities. The thick black line denotes the potential at the surface.}
	\label{Phi inside neutron star}
\end{figure}

\begin{figure}[h]
	\centering
	{\includegraphics[width=0.5\textwidth]{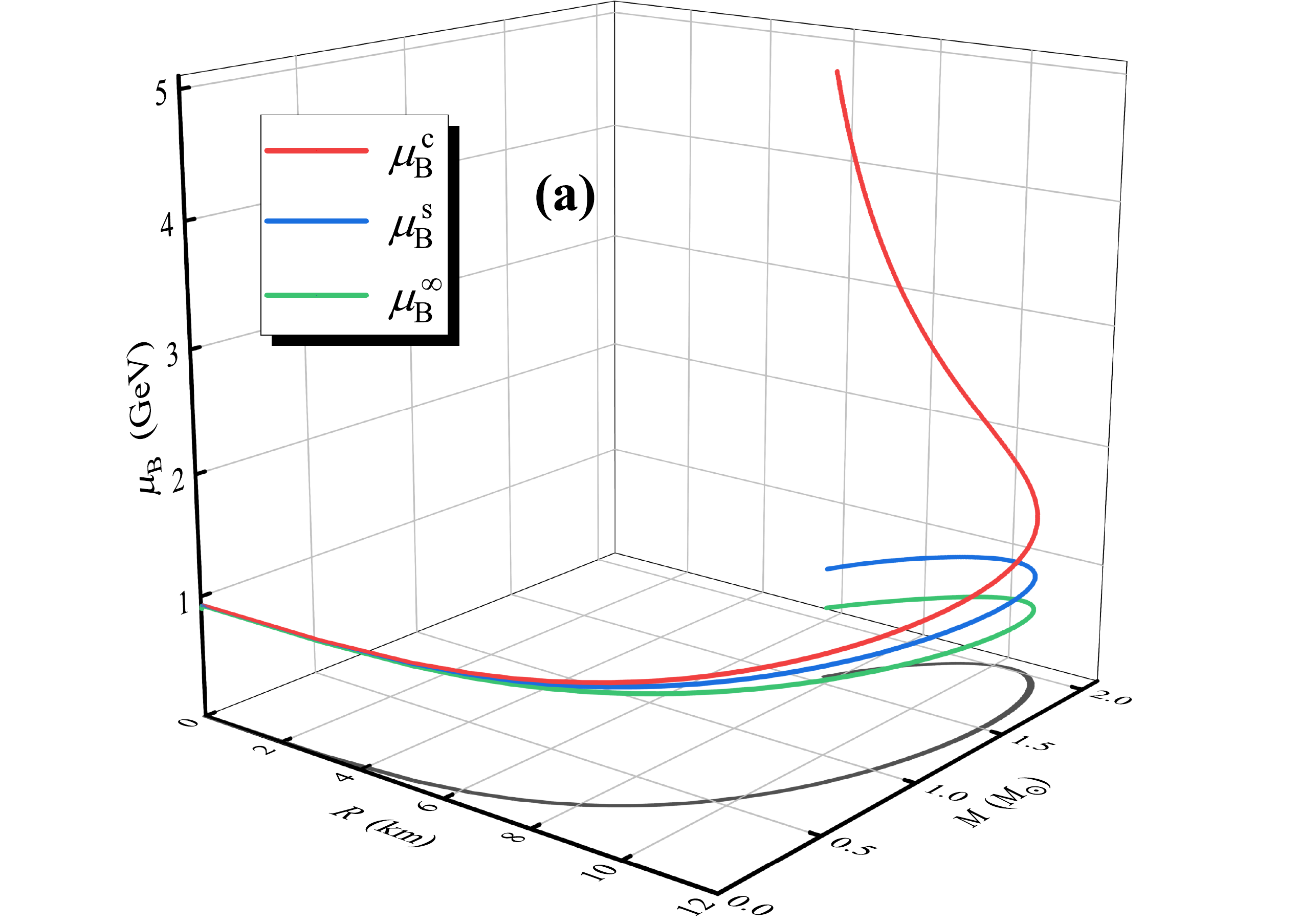}}
	{\includegraphics[width=0.4\textwidth]{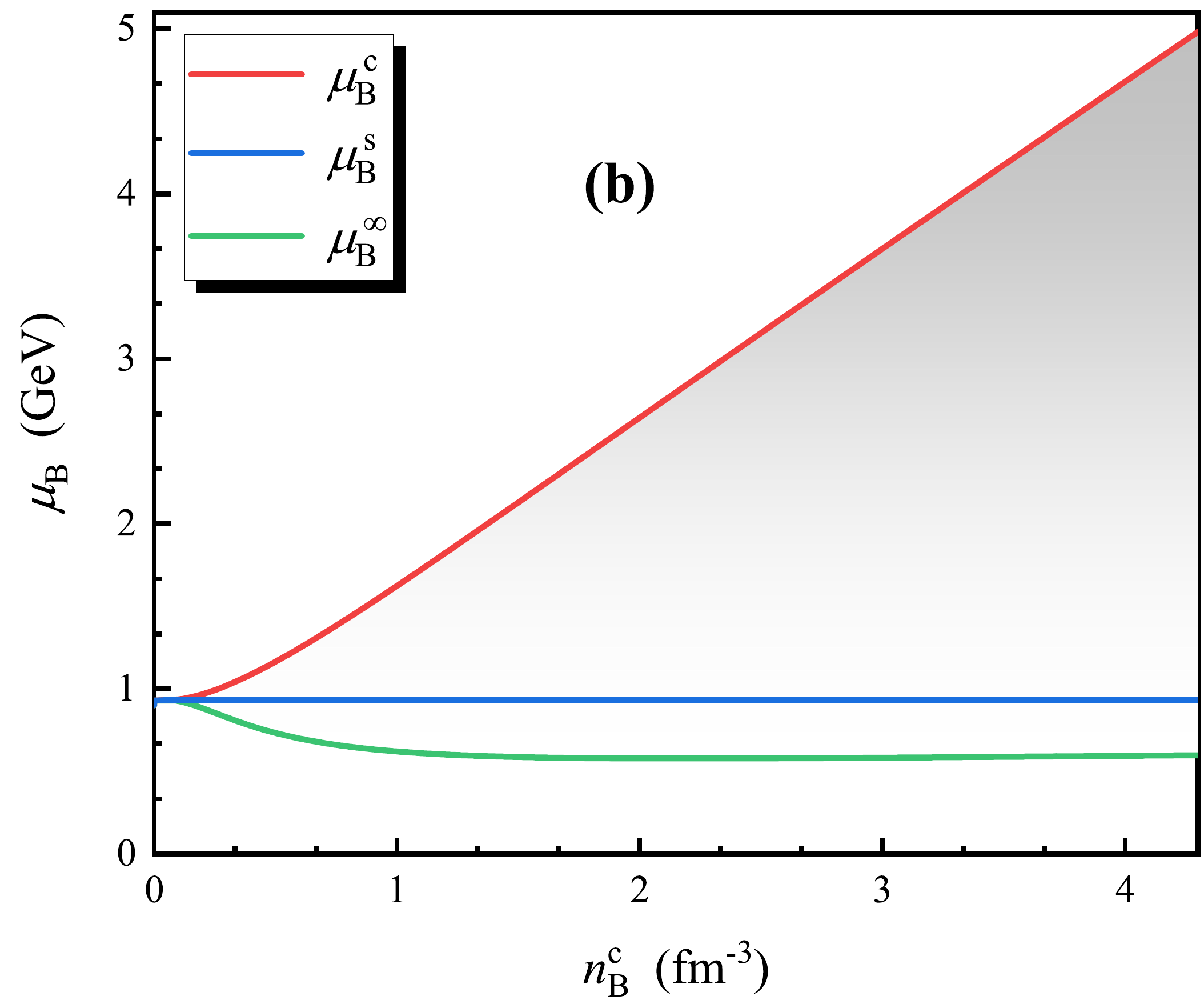}}
	\caption{The baryon chemical potentials $\mu_{\rm B}^{\rm c}$ at the core, $\mu_{\rm B}^{\rm s}$ at the surface, and $\mu_{\rm B}^\infty$ at infinity.
	(a) The three chemical potentials corresponding to the neutron stars on the mass-radius curve. (b) The three chemical potentials as functions of the central baryon density.}
	\label{chemical potential}
\end{figure}

With the known profile of the baryon chemical potential $\mu_{\rm B}(r)$ computed from the TOV equation (\ref{TOV3}),  the profile of the gravitational potential 
$\Phi(r)$ in the interior of neutron stars can be determined. The result is shown in Fig.~\ref{Phi inside neutron star}. To visualize how the gravitation enhances the baryon chemical potential in the interior of neutron stars, we display the baryon chemical potential $\mu_{\rm B}^\text{c}$ at the core, the baryon chemical potential $\mu_{\rm B}^\text{s}$ at the surface, and the redshifted one 
$\mu_{\rm B}^\infty$ in Fig.~\ref{chemical potential}. It is interesting to see that while a large central baryon density $n_{\rm B}^{\rm c}$ dramatically enhances the gravitational effect at the core, the gravitational potential at the surface, $\Phi_{\rm s}$, is almost a constant for sufficiently large central density ($\Phi_{\rm s}\simeq -0.5$ for $n_{\rm B}^{\rm c}>1{\rm fm}^{-3}$ in this model). As a result, the redshifted chemical potential $\mu_{\rm B}^\infty$ also reaches a platform at large central density, regardless of the large baryon chemical potential $\mu_{\rm B}^{\rm c}$ at the core. The difference between $\mu_{\rm B}^\infty$ and $\mu_{\rm B}^{\rm c}$ (or between $\mu_{\rm B}^{\rm s}$ and $\mu_{\rm B}^{\rm c}$) can be understood as an enhancement purely induced by the gravitational effect.

\section{Summary}\label{Sec5}
 
In summary, we have shown that to study the hydrostatic equilibrium of dense matter within the framework of general relativity and relativistic fluid dynamics, the EoS of dense matter 
should be that determined by many-body theories or experiments in flat spacetime so that it is compatible with local thermodynamic relations and conservation of energy and momentum. This is also protected by the equivalence principle as well.
We demonstrate this explicitly for the grand canonical EoS, which can be computed directly from field theoretical methods. As a by product, we demonstrate an alternative way to solve the TOV equation 
with the grand canonical EoS. In this approach, the enhancement of the baryon chemical potential inside the neutron star can be self-consistently regarded as a gravitational effect. This generalization 
may provide a way to extract the grand canonical EoS of dense matter via deep learning~\cite{Fujimoto:2021zas,Soma:2022qnv}.

\begin{acknowledgments}

This work was supported by the National Natural Science Foundation of China (Grant No. 11890712) the National Key R\&D Program (Grant No. 2018YFA0306503).

\end{acknowledgments}

\appendix
\section{EOS IN THE WALECKA MODEL}\label{App1}

\begin{figure}[H]
	\centering
	{\includegraphics[width=0.4\textwidth]{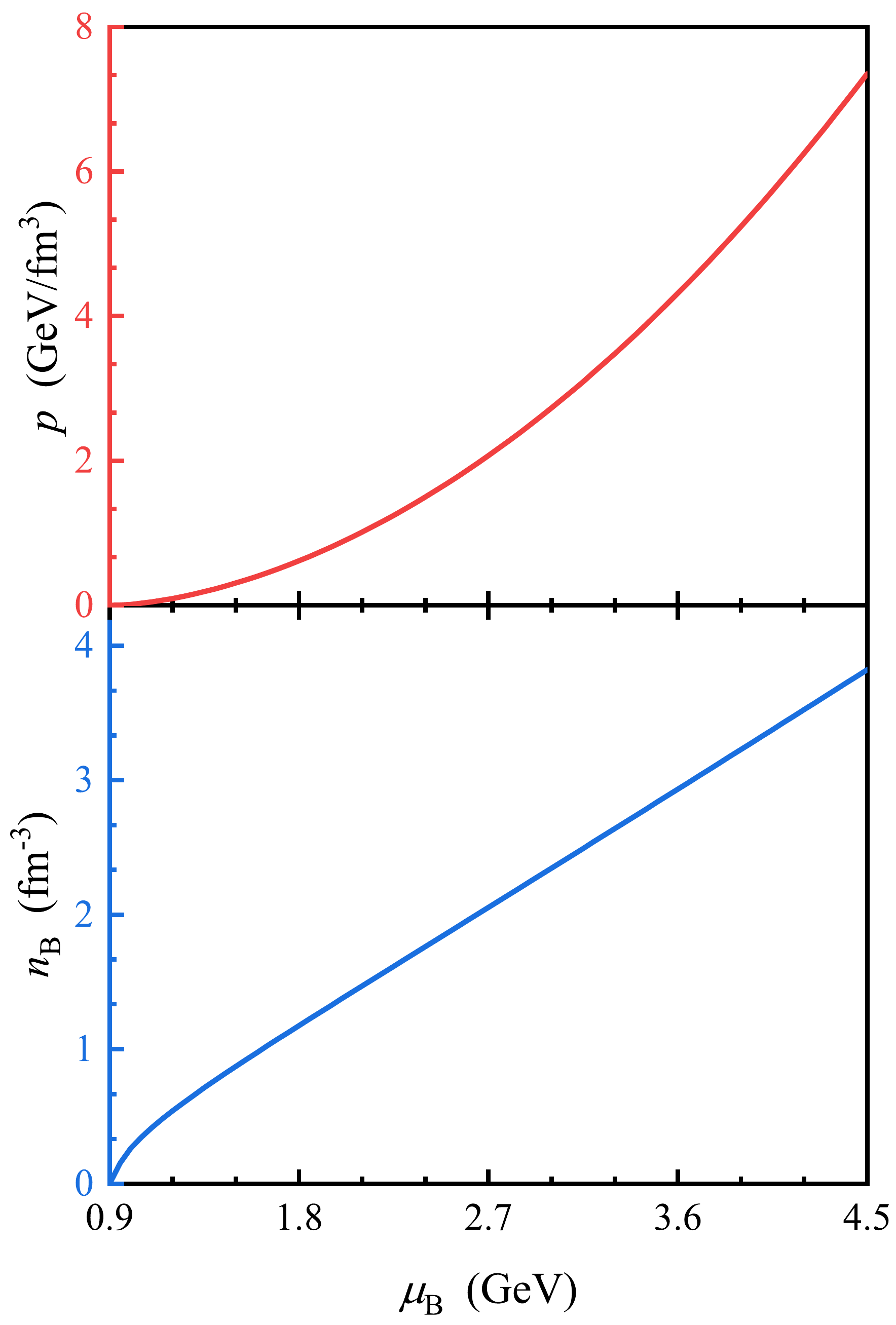}}
	\caption{The grand canonical EoS $p(\mu_{\rm B})$ and the corresponding baryon density $n_{\rm B}(\mu_{\rm B})$ of cold neutron star matter computed from the Walecka model.}
	\label{EoS-W}
\end{figure}

At zero temperature, the grand canonical EoS of dense matter in the Walecka model (without charge neutrality) can be evaluated as
\begin{eqnarray}
p\left(\mu_{\rm B},\mu_{\rm Q}\right)&=&\sum_{\rm N=n,p} p_0\left(\mu_{\rm N}^*, m_{\rm N}^*\right)+\sum_{l={\rm e},\mu}p_0\left(\mu_l,m_l\right) \nonumber\\
&&-\frac{1}{2} m_{\sigma}^{2} \bar{\sigma}^{2}-U(\bar{\sigma}) +\frac{1}{2} m_{\omega}^{2} \bar{\omega}_{0}^{2},
\end{eqnarray}
where the function $p_0(\mu,m)$ is defined as
\begin{eqnarray}
p_0\left(\mu,m\right)&=&\frac{1}{24 \pi^{2}}\Bigg[|\mu|\left(2\mu^2-5{m}^2\right) \sqrt{\mu^ 2-{m}^{2}}\nonumber\\
&+& 3 {m}^{4}~\text{arccosh}\left(\frac{|\mu|}{m}\right)\Bigg]\Theta(|\mu|-m).
\end{eqnarray}
The effective masses $m_{\rm N}^*$ and chemical potentials $\mu_{\rm N}^*$ are defined as
\begin{equation}
m_{\rm N}^*=m_{\rm N}-g_{\sigma}\bar{\sigma},\quad\mu_{\rm N}^*=\mu_{\rm N}-g_{\omega} \bar{\omega}_{0}.
\end{equation}
Note that the radiative correction from the vacuum contribution has been neglected. The meson condensates $\bar{\sigma}$ and $\bar{\omega}_0$ are determined by the following gap equations,
\begin{eqnarray}
&&m_{\omega}^{2} \bar{\omega}_{0}-\sum_{\rm N=n,p} g_{\omega} n_0\left(\mu_{\rm N}^*, m_{\rm N}^*\right)=0,\nonumber\\
&&m_{\sigma}^{2} \bar{\sigma}+U^\prime(\bar{\sigma})-\sum_{\rm N=n,p} g_{\sigma} n_{\rm s}\left(\mu_{\rm N}^*, m_{\rm N}^*\right)=0,
\end{eqnarray}
which minimize the effective potential ${\cal V}_{\rm eff}(\sigma,\omega_0)$. Here the functions $n_0(\mu,m)$ and $n_{\rm s}(\mu,m)$ are defined as
\begin{eqnarray}
&&n_0\left(\mu,m\right)=\frac{\left(\mu^2-m^2\right)^{3/2}}{3\pi^2}\Theta(|\mu|-m){\rm sgn}(\mu),\nonumber\\
&&n_{\rm s}\left(\mu,m\right)=\frac{m}{2\pi^2}\left[|\mu|\sqrt{\mu^2-m^2}-m^2\text{arccosh}\left(\frac{|\mu|}{m}\right)\right].\nonumber\\
\end{eqnarray}
The energy density $\varepsilon$ can be evaluated as
\begin{eqnarray}
\varepsilon\left(\mu_{\rm B},\mu_{\rm Q}\right)&=&\sum_{\rm N=n,p} \varepsilon_0\left(\mu_{\rm N}^*, m_{\rm N}^*\right)+\sum_{l={\rm e},\mu}\varepsilon_0\left(\mu_l,m_l\right) \nonumber\\
&&+\frac{1}{2} m_{\sigma}^{2} \bar{\sigma}^{2}+U(\bar{\sigma}) +\frac{1}{2} m_{\omega}^{2} \bar{\omega}_{0}^{2},
\end{eqnarray}
where the function $\varepsilon_0(\mu,m)$ reads
\begin{eqnarray}
\varepsilon_0\left(\mu,m\right)&=&\frac{1}{8\pi^2}\Bigg[|\mu|\left(2\mu^2-m^2\right)\sqrt{\mu^2-m^2}\nonumber\\
&-&m^4\text{arccosh}\left(\frac{|\mu|}{m}\right)\Bigg]\Theta(|\mu|-m).
\end{eqnarray}
The baryon density $n_{\rm B}$ and the electric charge density $n_{\rm Q}$ are given by
\begin{eqnarray}
n_{\rm B}\left(\mu_{\rm B},\mu_{\rm Q}\right)&=&\sum_{\rm N=n,p} n_0\left(\mu_{\rm N}^*, m_{\rm N}^*\right),\nonumber\\
n_{\rm Q}\left(\mu_{\rm B},\mu_{\rm Q}\right)&=& n_0\left(\mu_{\rm p}^*, m_{\rm p}^*\right)+\sum_{l={\rm e},\mu} n_0\left(\mu_{\rm Q}, m_{l}\right).
\end{eqnarray}
For neutron star matter, we should impose electric charge neutrality, 
\begin{eqnarray}
n_{\rm Q}\left(\mu_{\rm B},\mu_{\rm Q}\right)=0.
\end{eqnarray}
Thus the electric chemical potential is not an independent thermodynamic variable and should
be solved as $\mu_{\rm Q}=\mu_{\rm Q}(\mu_{\rm B})$. The grand canonical EoS of neutron star matter is the relation between the pressure and the baryon chemical potential, $p=p(\mu_{\rm B})$. The pressure $p(\mu_{\rm B})$ and the baryon density $n_{\rm B}(\mu_{\rm B})$ can be numerically evaluated for given model parameters, as shown in Fig.~\ref{EoS-W}. In the calculation, the model parameters are set as follows. The particle masses are taken as $m_{\rm n}=m_{\rm p}=m_{\rm N}=939$MeV, $m_\sigma=550$MeV, $m_\omega=783$MeV, $m_{\rm e}=0.511$MeV, and $m_\mu=105.66$MeV. The coupling constants are chosen as $g_\sigma=8.685$, $g_\omega=8.646$, $b=7.950\times10^{-3}$, and $c=6.952\times10^{-4}$ to fit the empirically known properties of nuclear matter~\cite{kapusta_gale_2006}.
\bibliographystyle{apsrev4-2}
\bibliography{Ref}
\end{document}